%% file: main.tex
\documentclass{article}

\usepackage{microtype}
\usepackage{graphicx}
\usepackage{subcaption}
\usepackage{booktabs} 

\usepackage{hyperref}

\usepackage[accepted]{icml2026}

\usepackage{amsmath}
\usepackage{amssymb}
\usepackage{mathtools}
\usepackage{amsthm}

\usepackage[capitalize,noabbrev]{cleveref}

\theoremstyle{plain}

\theoremstyle{definition}

\theoremstyle{remark}

\usepackage[disable,textsize=tiny]{todonotes}

\icmltitlerunning{Recency/Frequency Adaptive KV Caching for Large Language Model Serving}

\begin{document}

\twocolumn[
  \icmltitle{Recency/Frequency Adaptive KV Caching for Large Language Model Serving}



  \icmlsetsymbol{equal}{*}

  \begin{icmlauthorlist}
    \icmlauthor{Yang Shen}{JHU}
    \icmlauthor{Meghana Madhyastha$^\dagger$}{Parasail}
    \icmlauthor{Robert Underwood}{ANL}
    \icmlauthor{Bogdan Nicolae}{ANL}
    \icmlauthor{Randal Burns}{JHU}
  \end{icmlauthorlist}

  \icmlaffiliation{JHU}{Department of Computer Science, Johns Hopkins University, Baltimore, USA}
  \icmlaffiliation{ANL}{Argonne National Laboratory, Lemont, USA}
  \icmlaffiliation{Parasail}{Parasail, Inc., San Mateo, USA}

  \icmlcorrespondingauthor{Yang Shen}{yaangysshen@gmail.com}
  \icmlcorrespondingauthor{Randal Burns}{randal@cs.jhu.edu}

  \icmlkeywords{LLM Serving, KV Cache, Workload-Aware Caching}

  \vskip 0.3in
]



\printAffiliationsAndNotice{$^\dagger$ Work done while at Johns Hopkins University.}



\input{sec/0_abstract}

\input{sec/1_intro}
\input{sec/2_background}
\input{sec/3_method}
\input{sec/4_experiments}
\input{sec/5_conclusion}
\input{sec/6_future}
\input{sec/7_misc}


\bibliography{example_paper}
\bibliographystyle{icml2026}




\end{document}

%% file: sec/0_abstract.tex
\begin{abstract}
Key-value (KV) caching is a powerful technique for accelerating large language model inference and generation. Inference workloads are large and diverse, which makes them difficult to cache effectively. Existing cache management strategies adopt the least-recently-used policy for evicting cache blocks. However, LRU leads to  multiple unrelated workloads flushing each other's caches. To address this, we integrate adaptive caching that dynamically allocates cache space between recently and frequently occurring KV blocks. Evaluations show that it improves the KV cache hit rate by up to 10.8\% and reduces time to first token by up to 12.6\% over naive vLLM on synthetic document question answering workloads, and 2.1\% and 2.0\% respectively on real-world conversation workloads. The method generalizes well to batch inference and demonstrates clear interpretability while effectively accommodating diverse workloads. Our open-source implementation is available at \url{https://github.com/Y-aang/vllm-ARC}.

\end{abstract}

%% file: sec/1_intro.tex
\section{Introduction}
Large language models (LLMs) are widely deployed across a variety of applications, ranging from conversational agents to enterprise-level automation workflows. In production environments, the efficiency of inference and generation has emerged as a central concern~\cite{wan2023survey,zhou2024survey,zhen2025survey,miao2025survey}. The KV cache is a fundamental mechanism for reducing redundant computation during decoding~\cite{Orca,vLLM,Sglang}. However, as model sizes and context lengths continue to grow, the limited capacity of GPU memory has become a bottleneck for KV cache and raises new challenges for its effective management~\cite{H2o,Ada-kv,jiang2025towards}. 

Prefix caching and reuse between requests significantly reduces computation, and its hit rate directly affects both system throughput and latency~\cite{lmcache}. In many scenarios, different requests share a common prefix. For example, in multi-turn dialogue systems, a user’s historical conversation appears as a prefix in subsequent requests~\cite{Vicuna}. In question answering, certain documents or passages may act as hotspots frequently queried by different users~\cite{Wikiqa,QuALITY}. These observations suggest that designing efficient KV cache management strategies that capture such reuse behaviors between requests remains an open and impactful problem. 

However, most existing LLM serving frameworks, such as vLLM~\cite{vLLM}, primarily adopt simple least-recently-used (LRU) eviction strategies. Meanwhile, real workloads often exhibit more complex access characteristics: some present strong locality patterns, while others demonstrate a mixture of both recency and frequency properties. These mixed patterns in LLM serving are not fully captured by purely recency-driven eviction policies, motivating the exploration of more expressive strategies~\cite{Wild}. 

Therefore, we explore dynamic hybrid recency-frequency strategies as a foundation for adaptive KV cache management. We integrate the Adaptive Replacement Cache (ARC) \cite{ARC} into the LLM serving framework (vLLM). ARC implements a two-level split cache that separates recent and frequent requests; workload adaptive allocation of cache space between the two caches; and, ghost-caching to track misses that fall outside of the cache but could have been hits if cache space was allocated differently. 
We evaluate optimizations under diverse workload settings. Our study covers a long-context question answering (QA) workload derived from document QA datasets and a trace-driven conversational workload  from a large-scale LLM service. On the document QA workload, we observe up to a 10.8\% improvement in hit rate and a 12.6\% reduction in time to first token (TTFT); on the conversation workload, we observe up to a 2.1\% improvement in hit rate and a 2.0\% reduction in TTFT. These results indicate the potential of adaptive KV cache eviction and provide insights for future LLM serving systems. 

%% file: sec/2_background.tex
\section{Background and Related Work}

\noindent{\bf LLM Serving.}
The LLM inference process is auto-regressive, consisting of two stages: the prefill stage and the decoding stage~\cite{Flashattention,DistServe}. During the prefill stage, the model processes the entire input prompt in a single forward computation to obtain its hidden representations; during the decoding stage, the model generates new tokens one at a time conditioned on all preceding context.

To improve inference efficiency, modern LLM serving systems employ prefix caching and paged attention~\cite{vLLM,Sglang,Mooncake,CachedAttention,Preble}. Prefix caching reduces redundant computation by reusing key and value of self-attention for a shared prefix. It reuses the prefix within the same request during the decoding process, and also reuses the shared prefix across requests during the prefill process. Paged attention organizes the KV cache into fixed-size KV blocks and manages them through a paging-style allocation scheme~\cite{vLLM}, which avoids the need for contiguous memory, mitigates fragmentation, and improves efficiency.

For cache replacement, most systems~\cite{vLLM,Sglang} adopt recency-based policies such as LRU. When the cache becomes full, the system evicts the least-recently-used units (e.g., KV blocks) to make room for new requests. Marconi~\cite{Marconi} further explores FLOP-aware eviction over LRU for state-space models. RAGCache~\cite{Ragcache} designs a prefix-aware Greedy-Dual-Size-Frequency replacement policy for retrieval-augmented generation. MoonCake~\cite{Mooncake} introduces a hierarchical KV cache but focuses on prefill-decode disaggregation. Learning-based approaches~\cite{lpc} learn a continuation predictor to evict prefixes, while works like~\cite{Wild} propose workload-aware caching focusing on cloud use cases.



\noindent{\bf Applications.}
LLMs are now widely deployed in various scenarios, including document QA and multi-turn conversation. In document QA~\cite{Wikiqa,QuALITY}, the input typically consists of a long document paired with a question, requiring the model to process the entire document before producing an answer. In multi-turn conversation~\cite{Vicuna}, the dialog history grows with each turn, forming an increasingly large context that the model must repeatedly incorporate during generation.

Prefix caching and reuse are useful in those cases to improve inference efficiency. In document QA, the long document prefix can be reused through KV cache across requests. In multi-turn conversation, the accumulated dialogue history can be directly retrieved from the cache in each turn, reducing the cost of repeatedly reprocessing the entire context. 

%% file: sec/3_method.tex
\section{Method}

\begin{figure*}[t]
    \centering
    \includegraphics[width=1.0\textwidth]{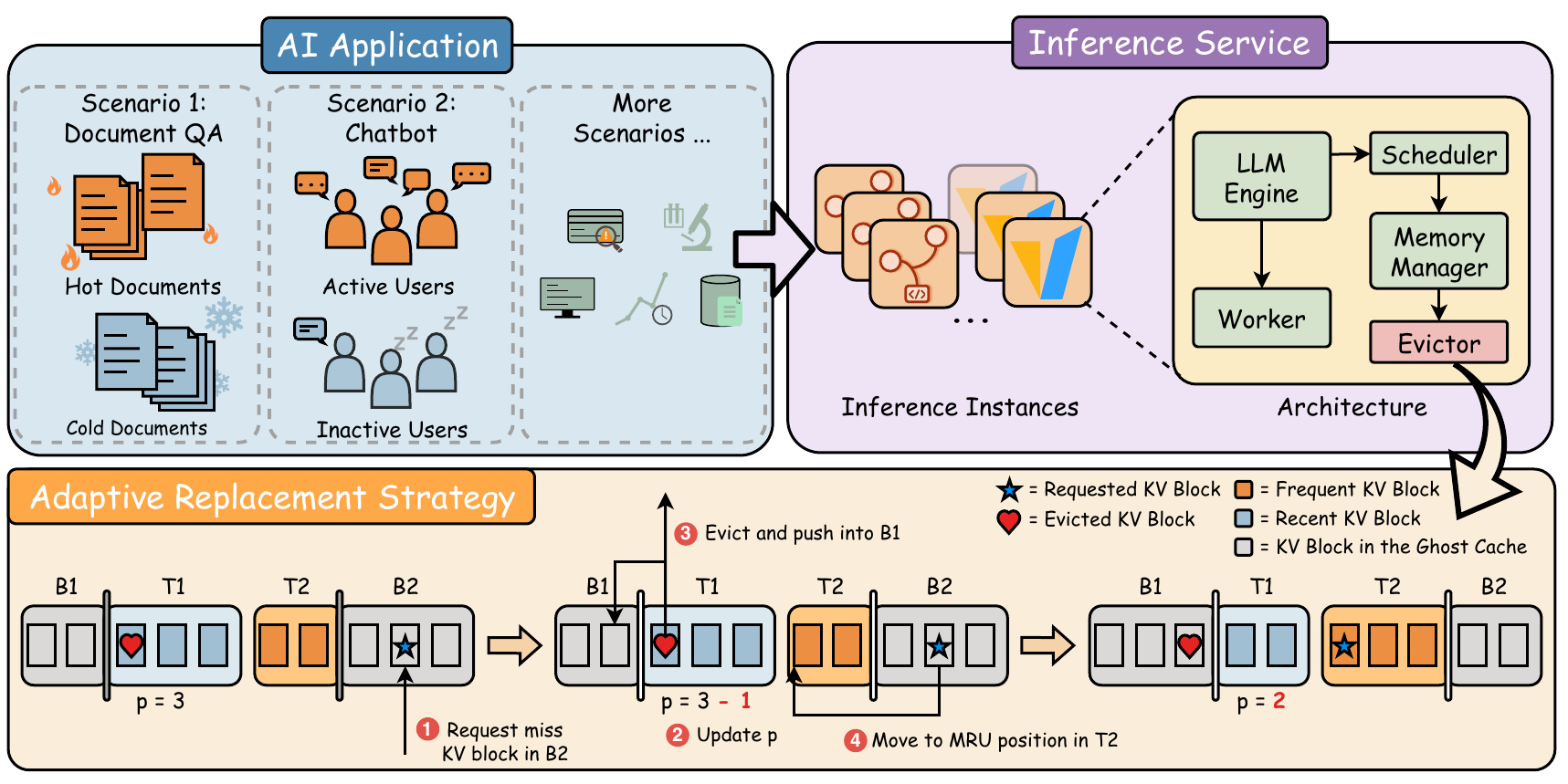}
    \caption{Overview of adaptive KV caching for LLM serving.}
    \label{fig:diagram}
\end{figure*}

\begin{figure}[t]
    \centering
    \includegraphics[width=1.0\linewidth]{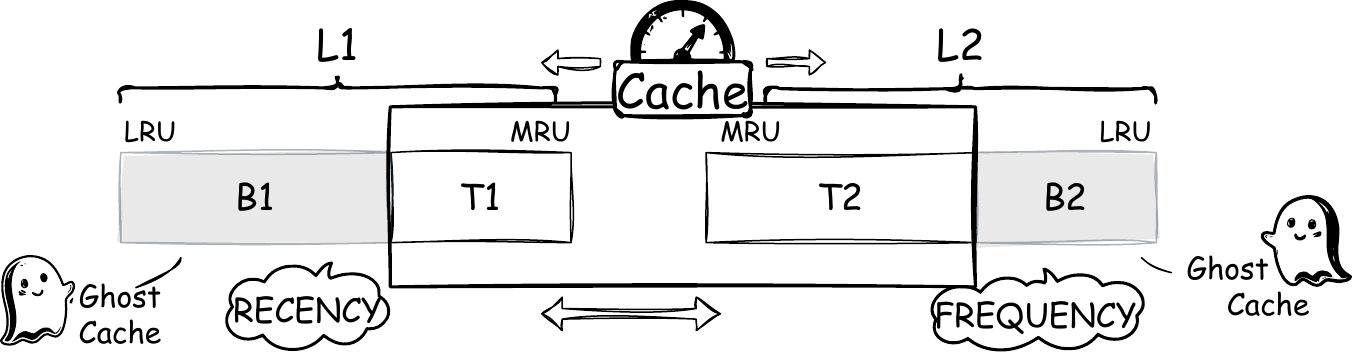}
    \caption{Structure of the Adaptive Replacement Cache. The cache consists of a recency queue (L1) and a frequency queue (L2) that are split into physical caches T1 and T2 and ghost caches B1 and B2 that track metadata. Requests that hit in the ghost caches are used to adapt the length of the T1 and T2 queues.}
    \label{fig:ARC}
\end{figure}

Modern LLM serving systems face large and dynamic workloads that exhibit a mixture of access patterns beyond simple locality. For instance, in document QA scenarios, specific documents often act as ``hotspots'' that are frequently queried by different users~\cite{Zipf_1,Zipf_2}. In multi-turn chatbot scenarios, sessions' temporal locality and varying user activity levels lead to complex and interleaved access patterns. These mixed patterns are not fully captured by purely recency-driven eviction policies. Motivated by this, we investigate hybrid eviction policies that dynamically partition cache space, allowing the system to adaptively fit diverse and shifting workloads.

Figure~\ref{fig:diagram} illustrates the overall architecture of the adaptive KV caching system for LLM serving. Requests from applications, such as document QA characterized by hot documents or chatbots characterized by active users, are handled by the inference service. The service routes requests to instances and the GPUs that process requests. Our technique addresses the management of the GPU caches, modifying the evictor module of a vLLM serving architecture. This applies to complex serving frameworks, including collaborative caching across multiple instances or multiple GPUs~\cite{Mooncake,lmcache}, and inference services that co-schedule or route requests based on knowledge of existing cache contents~\cite{Preble,cao2025locality}.
In these cases, there is a module that does per-GPU memory management and eviction. Our memory manager utilizes a specialized evictor based on the Adaptive Replacement Caching~\cite{ARC} to dynamically manage KV blocks based on real-time access patterns.

Figure~\ref{fig:ARC} illustrates the structure of ARC. This strategy automatically balances cache space between recency and frequency through two key features:

\begin{itemize}
    \item \textbf{Recency-Frequency Dual-Queue:} The cache is logically split into a low-frequency queue (L1 = T1 + B1) for items accessed once and a high-frequency queue (L2 = T2 +B2) for items accessed multiple times. Items in the low-frequency queue are promoted to the high-frequency queue upon a hit.
    \vspace{-1pt}
    \item \textbf{Dynamic Cache Partitioning with Ghost Caches:} ARC divides L1 and L2 into physical caches T1 and T2 that refer to KV blocks in memory and ghost caches B1 and B2 that extend the MRU/LRU history of T1 and T2. They store only the metadata for evicted KV blocks of each queue. Hits in the ghost cache serve as feedback signals to increase the size of the corresponding queue and dynamically adjust the partition of memory between two queues (T1 and T2).
\end{itemize}

The lower pane of Figure~\ref{fig:diagram} shows the operation of ARC. A request for a KV block that appears in B2 is a miss. The block is no longer cached. But, it has been referenced before and could have been a hit in a larger cache. The LLM engine processes the miss, regenerating the block. ARC puts this KV block to the frequency cache T2, evicts the LRU block from T1.
The combination of T1 and T2 is of fixed size. Misses that hit in B1 indicate that a larger T1 cache would have resulted in a hit and the algorithm should invest more space in T1. ARC~\cite{ARC} gives full details of adaptation and learning rate for a page cache.

The design makes the cache flush-resistant during workload shifts. A large number of new, unseen requests will clear T1, but the contents of T2 remain in cache. It adapts to workload shifts more quickly than a static policy by growing T1 when there is reuse of recently referenced KV blocks. It adapts to hotspots by growing T2 when KV blocks are used more than twice. 

The caching strategy is parameter-free and integrates naturally with the LLM inference engine. The KV cache is structured into fixed-size KV blocks (e.g., 16 tokens), aligning with the paged attention mechanism widely adopted in modern serving systems. All eviction decisions of the cache strategy are performed at the KV block level.

%% file: sec/4_experiments.tex
\section{Evaluation}
\begin{figure*}[t]
    \centering
    \begin{subfigure}{0.43\textwidth}
        \centering
        \includegraphics[width=\linewidth]{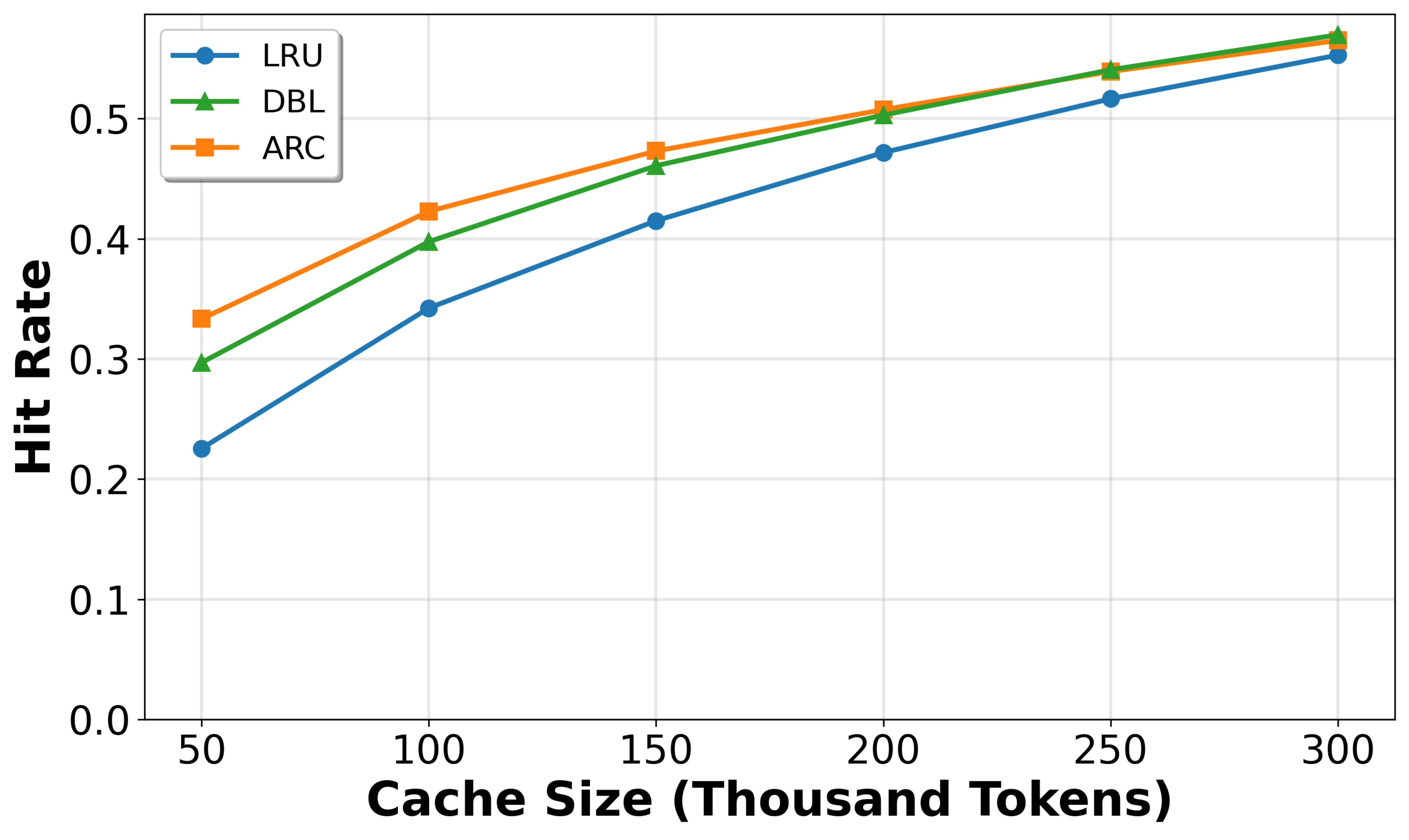}
        \caption{KV Cache Hit Rate}
        \label{fig:quality hit rate}
    \end{subfigure}
    \hspace{0.06\textwidth}
    \begin{subfigure}{0.43\textwidth}
        \centering
        \includegraphics[width=\linewidth]{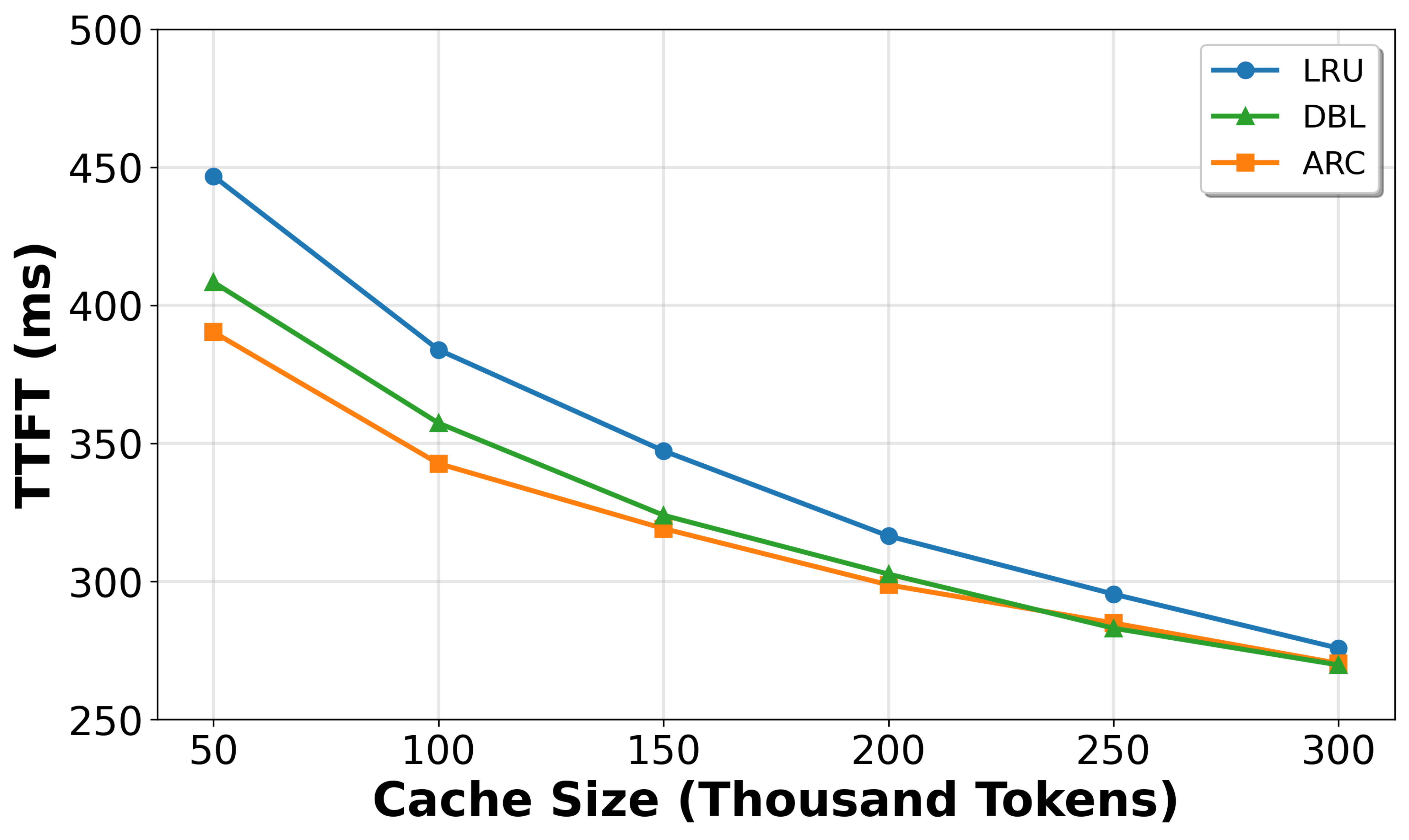}
        \caption{Average TTFT}
        \label{fig:quality ttft}
    \end{subfigure}
    \caption{
        Document QA workload on QuALITY that compares static two-queue replacement (DBL) and adaptive two-queue replacement (ARC) with LRU. \;
    }
    \label{fig:quality exp}
\end{figure*}

\begin{figure*}[t]
    \centering
    \begin{subfigure}{0.43\textwidth}
        \centering
        \includegraphics[width=\linewidth]{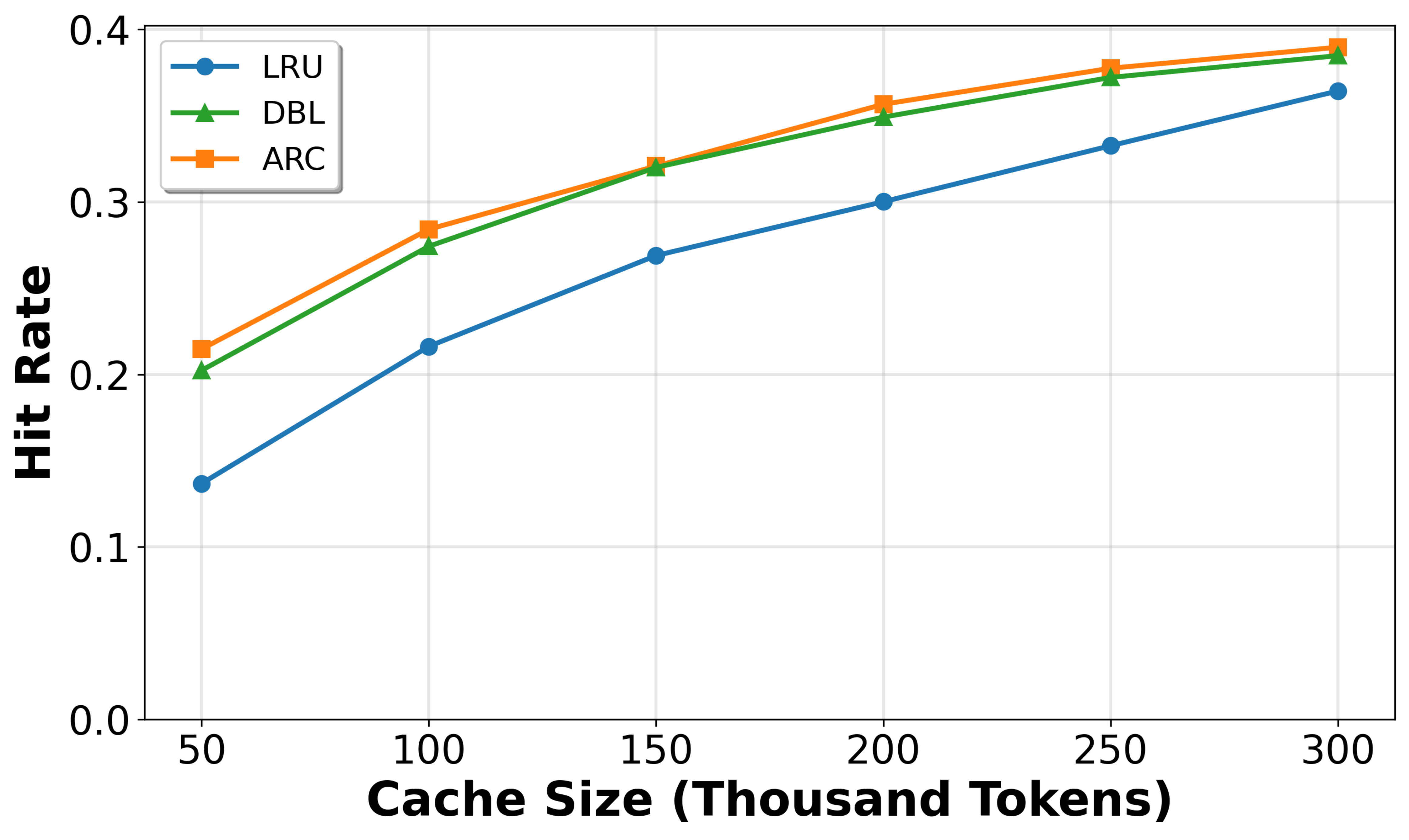}
        \caption{KV Cache Hit Rate}
        \label{fig:wikiqa hit rate}
    \end{subfigure}
    \hspace{0.06\textwidth}
    \begin{subfigure}{0.43\textwidth}
        \centering
        \includegraphics[width=\linewidth]{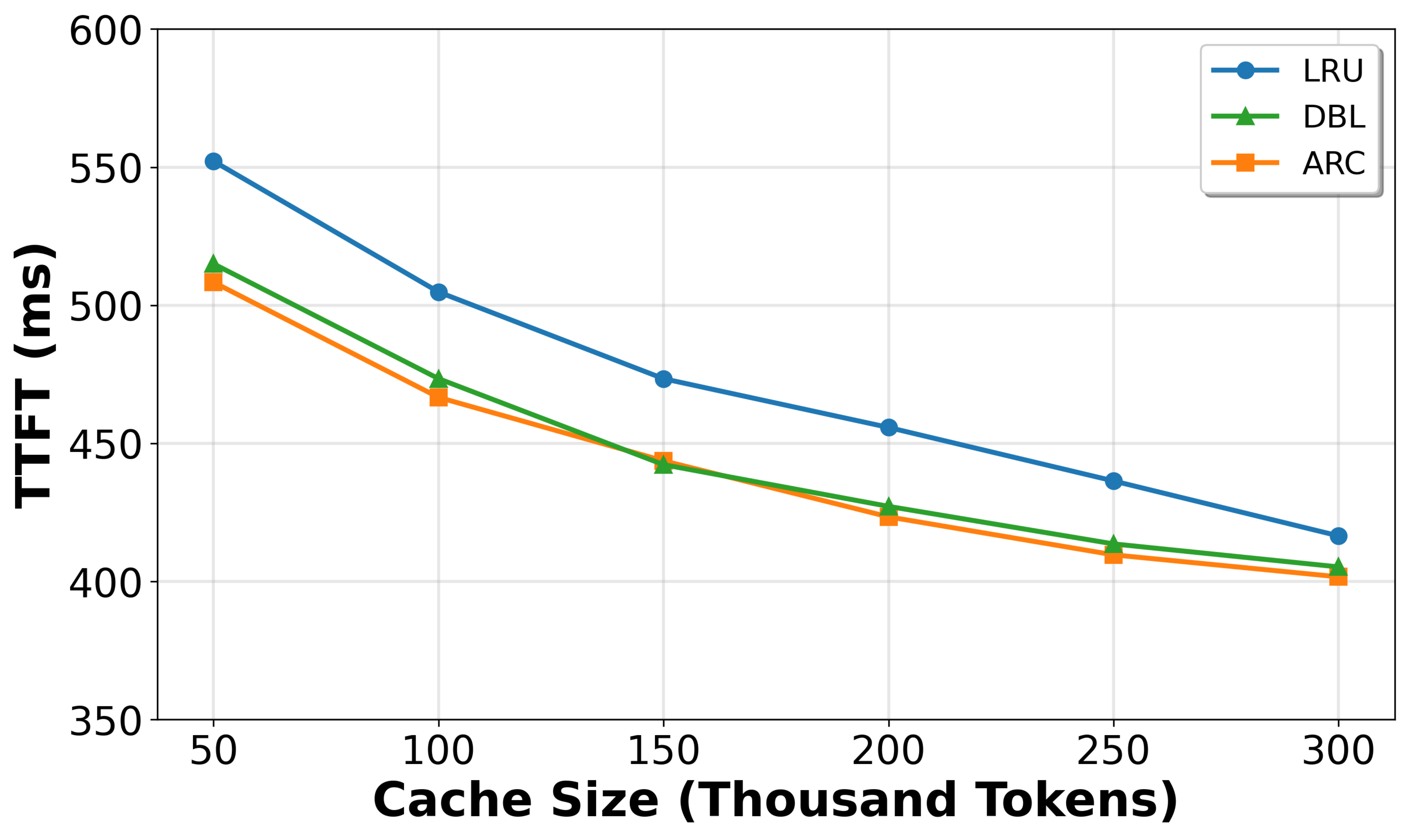}
        \caption{Average TTFT}
        \label{fig:wikiqa ttft}
    \end{subfigure}
    \caption{
        Document QA workload on WikiQA. \;
    }
    \label{fig:wikiqa exp}
\end{figure*}

We integrate the adaptive caching strategy into vLLM~\cite{vLLM} and evaluate performance for two document QA workloads and a multi-turn conversation trace-driven workload. Experiments measure the cache hit rate and correlate improvements to the performance metric TTFT during generation. We isolate the contributions of both aspects of ARC: two-queue caching and adaptive sizing. We demonstrate effectiveness across different batch sizes. Finally, we examine the sizes of the recency and frequency caches over time to demonstrate the importance of adaptation.


\noindent \textbf{Model and Hardware Configuration.} We evaluate all eviction strategies using Qwen3~\cite{Qwen3} with 14B parameters. Qwen3 is a popular open-source model, and 14B is a representative model size for modern LLM serving. All the main experiments are conducted on a single NVIDIA H100 PCIe GPU with 80 GB of memory on Lambda GPU Cloud.


\noindent \textbf{Inference Configuration.}  To isolate the effect of cache eviction strategies from other runtime factors, we standardize the inference behavior:

\begin{itemize}
    \item The offline inference mode is used. Except when noted, the batch size is fixed to 1, eliminating requests waiting on prior requests in the batch, ensuring that all TTFT variation reflects cache behavior.
    \item Each request generates exactly one token, removing variability from the decoding phase to prevent generating responses of different lengths. 
    \item The KV block size is set to 16 tokens, following the standard paged-attention layout used in modern LLM serving systems.
    \item Prefix caching is enabled, allowing shared prefixes across requests to be reused by the eviction strategies.
\end{itemize}





\subsection{Document Question Answering}\label{sec:DQA}
We evaluate cache eviction strategies using two widely-adopted long-context document QA datasets: QuALITY~\cite{QuALITY} and WikiQA~\cite{Wikiqa}. QuALITY is a long-document reading comprehension dataset. We apply HTML-stripped v1.0.1 from the official repository, which contains 381 unique long documents paired with 761 questions with passages over 5k tokens on average. WikiQA is an open-domain question-answering dataset built from Bing search queries and their linked Wikipedia pages. It includes 3,047 user questions mapped to their associated long passages, along with 29,258 question-sentence pairs. 

\begin{figure*}[t]
    \centering
    \begin{subfigure}{0.43\textwidth}
        \centering
        \includegraphics[width=\linewidth]{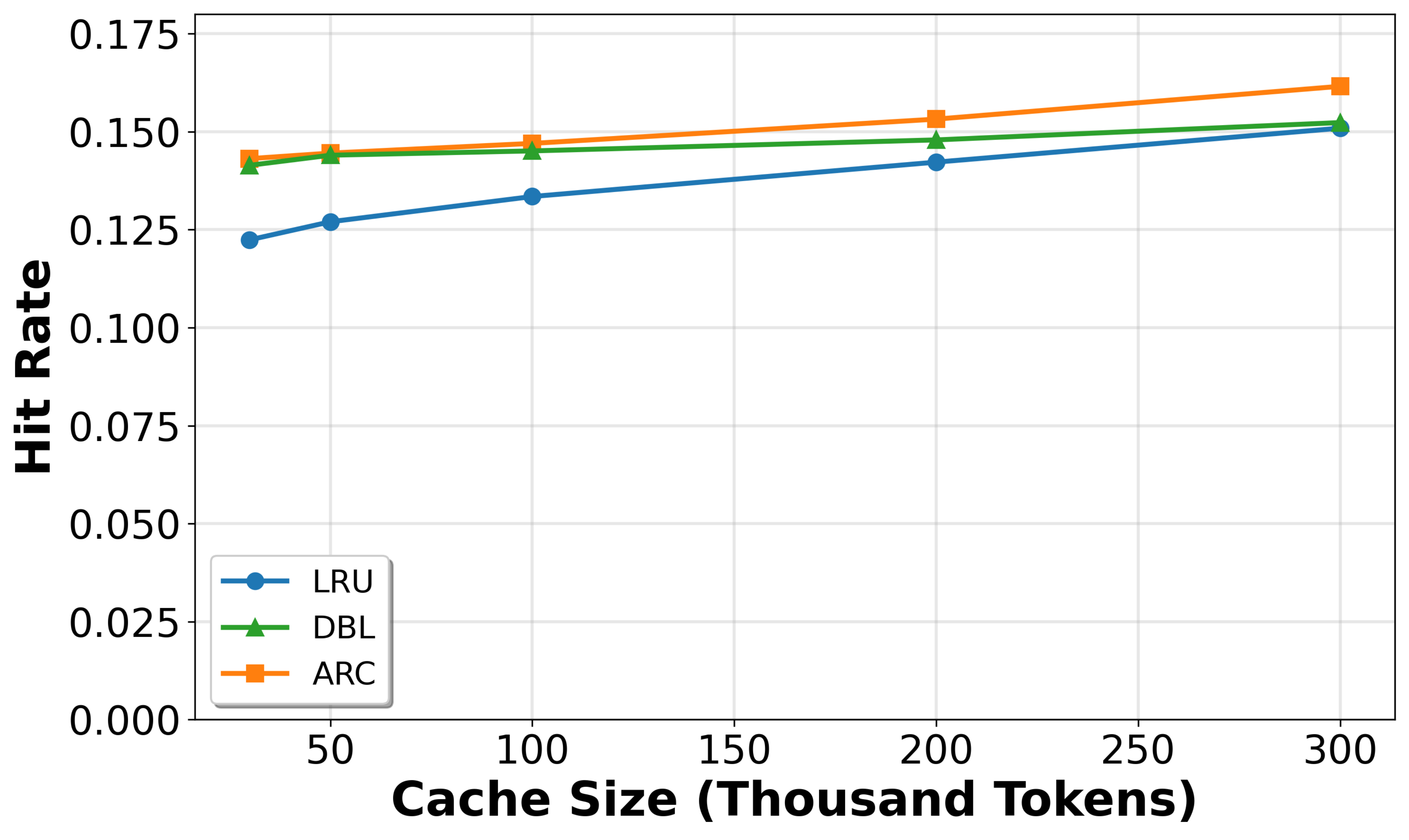}
        \caption{KV Cache Hit Rate}
        \label{fig:conversation hit rate}
    \end{subfigure}
    \hspace{0.06\textwidth}
    \begin{subfigure}{0.43\textwidth}
        \centering
        \includegraphics[width=\linewidth]{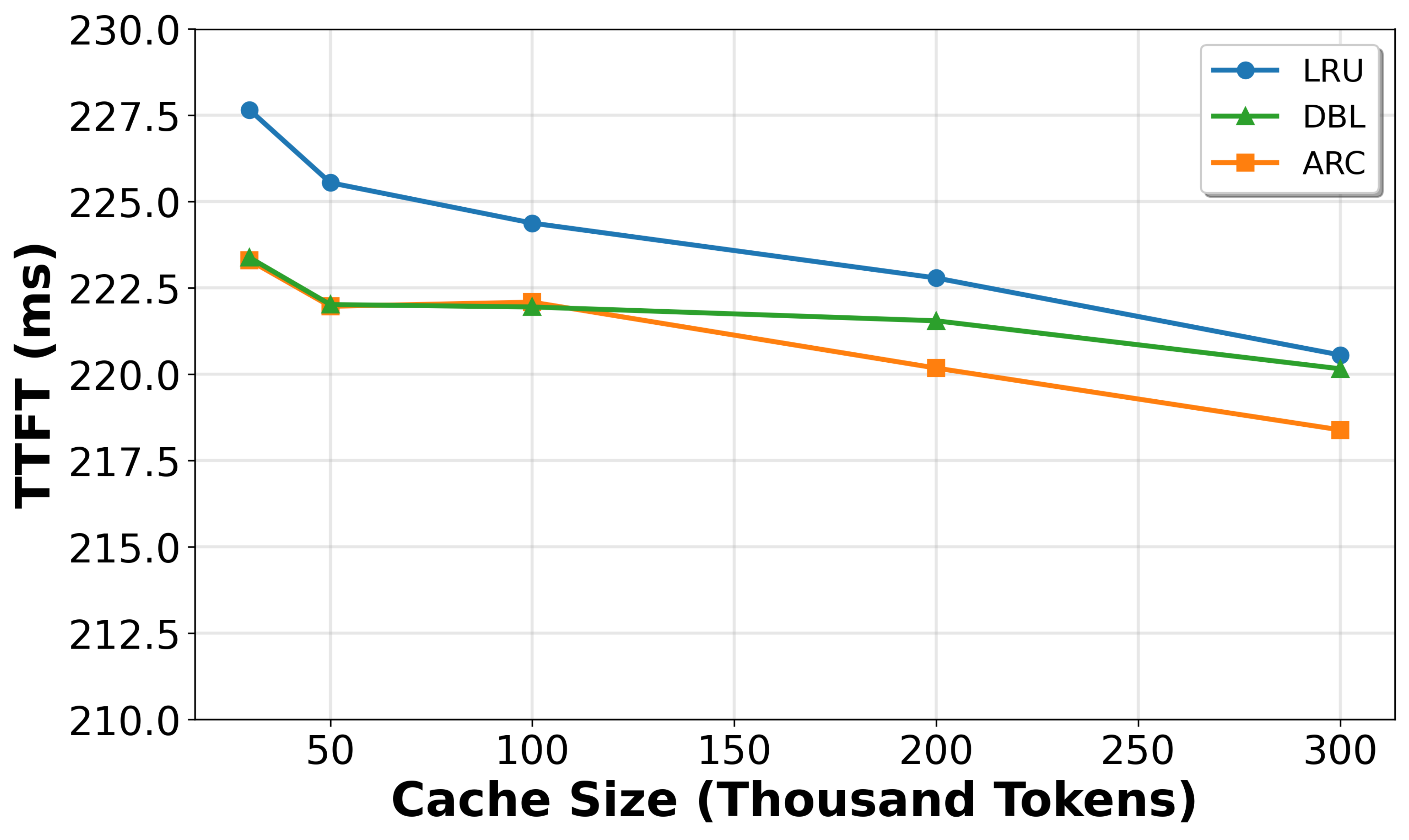}
        \caption{Average TTFT}
        \label{fig:conversation ttft}
    \end{subfigure}
    \caption{
        Performance comparison on the multi-turn conversation workload. \;
    }
    \label{fig:conversation exp}
\end{figure*}

These are question answer sets, not workloads. We know of no publicly available trace-driven QA workloads. We generate synthetic workloads from them using windowed Zipf sampling, which is widely used to generate workloads with realistic distribution for LLMs~\cite{Preble,burstgpt,flashinfer}.
 Specifically, we draw 3,072 requests in total and partition them into windows of 512 requests each. Within each window, requests are sampled according to a Zipf distribution (exponent = 1.0), while the underlying items are reshuffled across windows to introduce shifts in popularity.

 We evaluate and compare ARC with the vLLM LRU replacement algorithm. We also include {\bf DBL}, a variant of ARC that employs a fixed 1:1 ratio between recency and frequency queues, to isolate the contribution of recency/frequency separation from memory adaptation.

Figure~\ref{fig:quality exp} and \ref{fig:wikiqa exp} show that ARC improves the KV cache hit rate and that translates to an overall improvement in TTFT.  ARC increases the hit rate by 1.2\%-10.8\% on QuALITY and 2.5\%-7.8\% on WikiQA. The average TTFT under ARC decreases by 2.0\%-12.6\% and 3.6\%-7.9\% respectively. As the cache size grows, the improvement of the adaptive eviction policy diminishes. This is expected as larger caches capture a larger fraction of the workload and replacement decisions are less frequent and less important.
In these workloads, improvements from the adaptive properties of ARC less important and more pronounced at smaller cache sizes.

\subsection{Multi-Turn Conversation}
We use Trace A from the Qwen-Bailian Anonymous Dataset to evaluate the adaptive cache eviction under a multi-turn conversation scenario~\cite{Wild}. This trace contains a two-hour sample of anonymized KV cache requests served by a single Qwen deployment on Aliyun Bailian, one of the world’s largest cloud providers. Trace A corresponds to a ChatGPT-style, consumer-facing service, capturing real user interactions at production scale. It contains 43,058 requests, and the average input length is around 2.3k tokens. It provides a practical view of multi-turn conversation scenario. To ensure stable evaluation, we warm up the cache with the first 20,000 requests and report hit rate and TTFT on the rest.


On a trace-driven conversational workload, ARC improves performance compared with the baseline LRU (Figure~\ref{fig:conversation exp}). ARC achieves a 1.1\%-2.1\% increase in hit rate, which translates into a 1.0\%-2.0\% reduction in average TTFT. In this case, the contribution from adaptive queue sizing grows as the cache grows in size and static two-queue caching (DBL) degrades for caches larger than 200K tokens. This represents workload shifts on a longer scale that arise only in larger caches (see Section \ref{sec:twoq}).

\subsection{Batch Inference}\label{sec:Batch}
\begin{figure}[t]
    \centering
    \includegraphics[width=\linewidth]{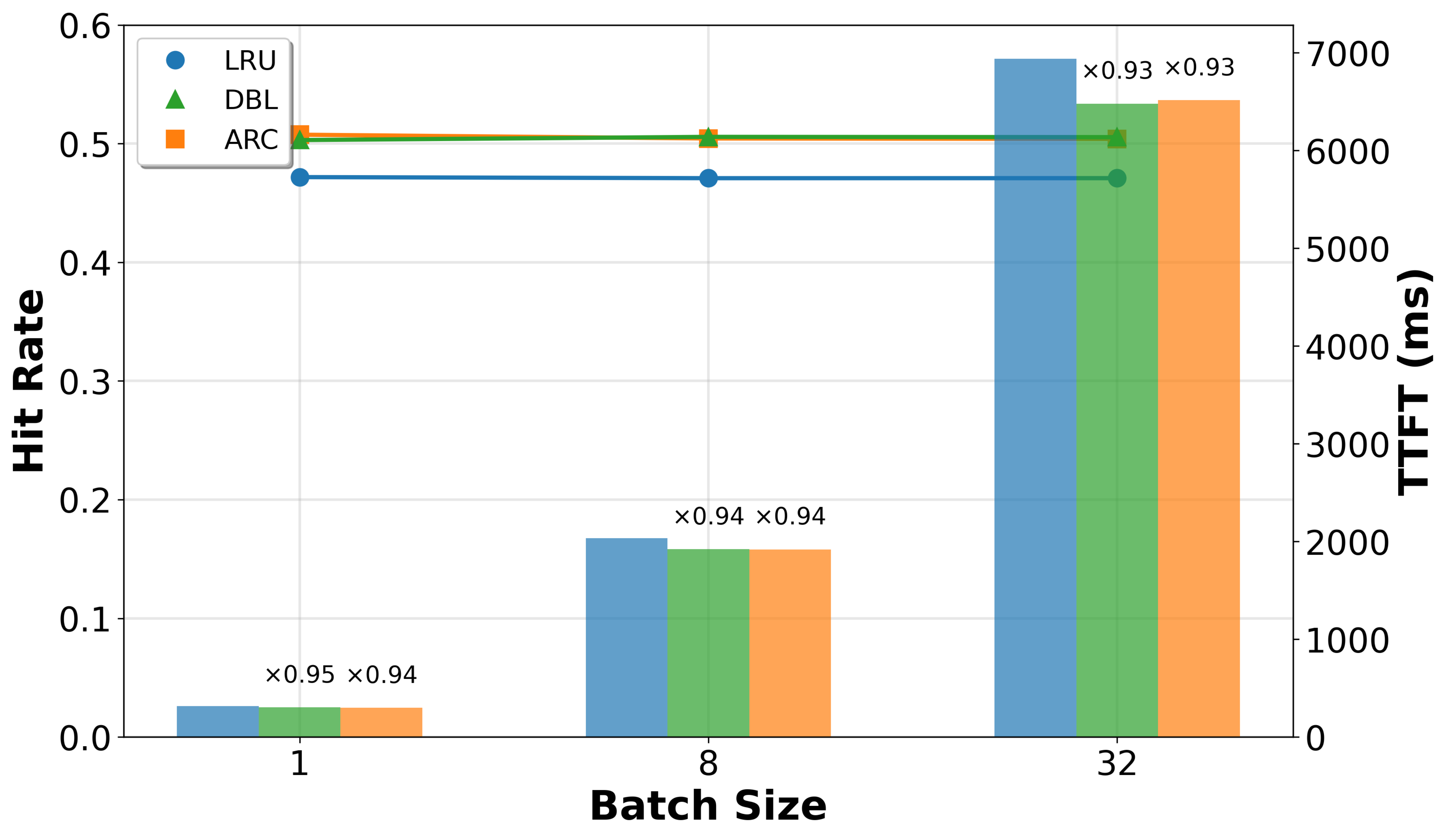}

    \caption{KV cache hit rate and average TTFT under different batch sizes on the QuALITY dataset. The cache size is 20k tokens.}
    \label{fig:BatchExp}
\end{figure}

To evaluate whether adaptive KV caching remains effective under batch inference, we conduct additional experiments using the same synthetic document QA workload as in Section~\ref{sec:DQA}, while varying the batch size. Batch inference is commonly adopted in offline serving and throughput-oriented scenarios, where multiple requests are processed simultaneously to maximize throughput. In this experiment, the cache size is set as 20 thousand tokens.

\begin{figure*}[t]
    \centering
    \includegraphics[width=\linewidth]{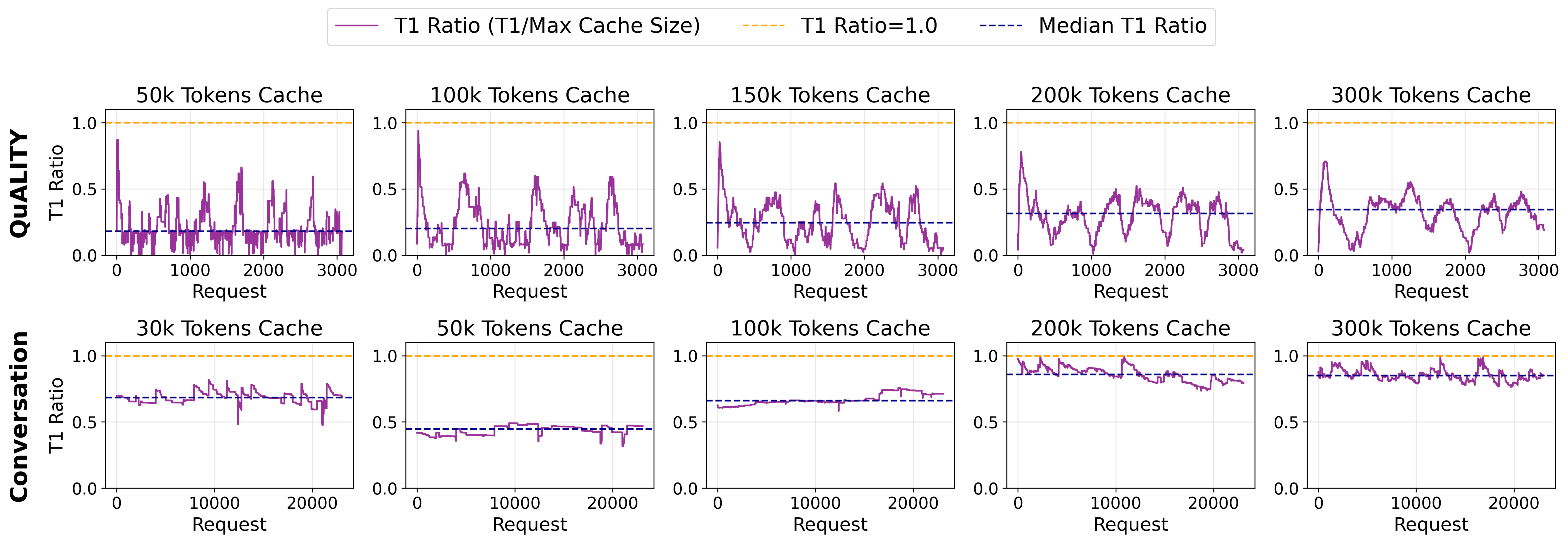}

    \caption{Dynamic evolution of the T1 Ratio across different workloads and cache capacities.}
    \label{fig:Ablation}
\end{figure*}

Figure~\ref{fig:BatchExp} reports the KV cache hit rate and average TTFT under different batch sizes on the QuALITY dataset. The cache hit rate remains stable as the batch size increases for all eviction strategies. This trend is expected, as the increase of batch size does not change much the prefix reuse pattern between requests. The only difference would be if the same prefix arose in the same batch.

The relative advantage of ARC persists as the batch size grows compared to LRU with a 5-7\% reduction in TTFT and benefits grow slightly as batch size increases. The increase in the time to first token as batch size increases reflects that the larger batch delays all requests. Throughput is higher on aggregate, but TTFT is reduced. 


\subsection{Impact of Adaptive Partitioning} \label{sec:twoq}

ARC's improvement comes in part from the dynamic allocation of memory between the two caches.
Examining the evolution of the T1 and T2 cache sizes over time provides insight into how adaptive memory allocation improves performance. It also characterizes workloads, revealing dynamics in reuse and workload shifts. 
Figure~\ref{fig:Ablation} shows the ratio of T1 (recency queue) and T2 (frequency queue) over the duration of the experiment. 

The conversation dataset reveals the operation of a cache under a trace-driven workload. As the cache increases in size, the balance of space is reserved for recency. This hints at a smaller working set maintained for frequency that is already captured at smaller cache sizes. The 30K cache size seems to be an outlier. Likely, 30K tokens is not large enough to store the working set and the cache is unable to populate T2. This reflects the poor hit rate seen at this cache size (Figure \ref{fig:conversation exp}). 

We also see that the cache sizes change more dramatically at 200K and 300K in contrast to the 100K token cache. ARC captures long-scale workload shifts that are not captured by smaller ARC caches or other methods. The adaptation leads to a divergence in performance between ARC and DBL (Figure \ref{fig:conversation exp}).

The QuALITY dataset shows the cache migrating between recency and frequency. At the start, the cache is empty and there are no frequency hits. The cache invests all space in the recency queue. Over many requests, the cache migrates between frequency and recency. This corresponds to 512-request windowed Zipf sampling used to generate the workload. Again, the median shifts to recency as the cache size increases.

In both workloads, the deviation of the T1 ratio from the 50\% baseline aligns with the performance gain, which shows that the adaptive strategy outperforms a static split.



%% file: sec/5_conclusion.tex
\section{Conclusion}
This paper presents adaptive KV caching for LLM serving, integrating hybrid recency-frequency strategies (ARC) into vLLM. By adaptively capturing prefix-reuse patterns beyond simple recency, these strategies improve cache hit rate and reduce TTFT across both synthetic document QA workloads and real multi-turn conversation traces. Our method generalizes naturally to batch inference and exhibits clear interpretability, demonstrating that adaptive KV cache management is a practical and effective way to enhance LLM serving performance.

%% file: sec/6_future.tex
\section{Future Work}
An extension of this work is to expand the scope of adaptability within the eviction framework. Beyond recency and frequency, other factors such as prompt-length-dependent latency savings and positional variance of reuse probability could be explored. Designing eviction strategies that explicitly balance length, recency, and frequency is a direction for further performance improvements.

Other directions include evaluating and designing caching strategies in more sophisticated serving architectures. While our design and evaluation are primarily based on offline inference, production systems often involve online inference, hierarchical CPU/Disk offloading, multi-GPU distributed setups, and prefill-decoding disaggregated architectures. Exploring diverse scenarios presents opportunities to further refine system design and optimize end-to-end performance.



%% file: sec/7_misc.tex
\section*{Impact Statement}
This paper presents work whose goal is to advance the field of machine learning and system for machine learning. There are many potential societal consequences of our work, none of which we feel must be specifically highlighted here.

\section*{Acknowledgements}
This material is based upon work supported by the U.S. Department of Energy (DOE), Office of Science, Office of Advanced Scientific Computing Research, under Contract DE-AC02-06CH11357 and by the National Science Foundation under Grant NSF OAC-02103874.
Any opinions, findings, and conclusions or recommendations expressed in this material are those of the author(s) and do not necessarily reflect the views of the National Science Foundation or Department of Energy.